\documentclass[reprint, superscriptaddress,
amsmath,amssymb,
aps,
prl,
 floatfix,
]{revtex4-1}

\usepackage[normalem]{ulem} 

\usepackage{natbib}
\usepackage[colorlinks=true,citecolor=blue,linkcolor=blue,urlcolor=blue]{hyperref}
\usepackage{prlfonts}
\usepackage{graphicx}
\usepackage{dcolumn}
\usepackage{bm}
\usepackage[dvipsnames]{xcolor}
\definecolor{red}{RGB}{250, 0, 0}
\usepackage{amsmath}
\usepackage{mathtools}
\usepackage{color,soul}

\begin{document}



\title{Enhanced One-Color-Two-Photon Resonant Ionization in Highly Charged Ions by Fine-Structure~Effects}

\author{Moto Togawa}
\email{moto.togawa@xfel.eu}
\affiliation{European XFEL, Holzkoppel 4, 22869 Schenefeld, Germany}%
\affiliation{Max-Planck-Institut f\"ur Kernphysik, Saupfercheckweg 1, 69117 Heidelberg, Germany}%

\author{Chunhai Lyu}
\email{chunhai.lyu@mpi-hd.mpg.de}
\affiliation{Max-Planck-Institut f\"ur Kernphysik, Saupfercheckweg 1, 69117 Heidelberg, Germany}%

\author{Chintan~Shah}
\affiliation{NASA Goddard Space Flight Center, 8800 Greenbelt Rd,
Greenbelt, MD 20771, USA}
\affiliation{Max-Planck-Institut f\"ur Kernphysik, Saupfercheckweg 1,
69117 Heidelberg, Germany}
\affiliation{Department of Physics and Astronomy, Johns Hopkins
University, Baltimore, Maryland 21218, USA}

\author{Marc Botz}
\affiliation{Max-Planck-Institut f\"ur Kernphysik, Saupfercheckweg 1, 69117 Heidelberg, Germany}%
\affiliation{Heidelberg Graduate School for Physics, Ruprecht-Karls-Universität Heidelberg,
Im Neuenheimer Feld 226, 69120 Heidelberg, Germany}%

\author{Joschka Goes}
\affiliation{Max-Planck-Institut f\"ur Kernphysik, Saupfercheckweg 1, 69117 Heidelberg, Germany}%

\author{Jonas Danisch}
\affiliation{Max-Planck-Institut f\"ur Kernphysik, Saupfercheckweg 1, 69117 Heidelberg, Germany}%

\author{Marleen~Maxton}
\affiliation{Max-Planck-Institut f\"ur Kernphysik, Saupfercheckweg 1, 69117 Heidelberg, Germany}%

\author{Kai~K\"obnick}
\affiliation{Max-Planck-Institut f\"ur Kernphysik, Saupfercheckweg 1, 69117 Heidelberg, Germany}%

\author{Filipe Grilo}%
\affiliation{%
 Laboratory of Instrumentation, Biomedical Engineering and Radiation Physics (LIBPhys-UNL), Department of Physics, NOVA School of Science and Technology, NOVA University Lisbon, 2829-516 Caparica, Portugal 
}%

\author{Pedro Amaro}%
\affiliation{%
 Laboratory of Instrumentation, Biomedical Engineering and Radiation Physics (LIBPhys-UNL), Department of Physics, NOVA School of Science and Technology, NOVA University Lisbon, 2829-516 Caparica, Portugal 
}%

\author{Katharina Kubicek}
\affiliation{European XFEL, Holzkoppel 4, 22869 Schenefeld, Germany}%
\affiliation{Department of Physics, Universität Hamburg, Notkestraße 9-11, 22607 Hamburg, Germany}

\author{Mohammed Sekkal}
\affiliation{European XFEL, Holzkoppel 4, 22869 Schenefeld, Germany}%
\affiliation{Department of Physics, Universität Hamburg, Notkestraße 9-11, 22607 Hamburg, Germany}

\author{Awad Mohamed}%
\affiliation{%
 ISM-CNR, Istituto di Struttura dei Materiali, LD2 Unit, 34149 Trieste, Italy
}%

\author{Rebecca Boll}
\affiliation{European XFEL, Holzkoppel 4, 22869 Schenefeld, Germany}%

\author{Alberto~De~Fanis}
\affiliation{European XFEL, Holzkoppel 4, 22869 Schenefeld, Germany}%

\author{Simon Dold}
\affiliation{European XFEL, Holzkoppel 4, 22869 Schenefeld, Germany}%

\author{Tommaso Mazza}
\affiliation{European XFEL, Holzkoppel 4, 22869 Schenefeld, Germany}%

\author{Jacobo Montano}
\affiliation{European XFEL, Holzkoppel 4, 22869 Schenefeld, Germany}%

\author{Nils Rennhack}
\affiliation{European XFEL, Holzkoppel 4, 22869 Schenefeld, Germany}%

\author{Björn Senfftleben}
\affiliation{European XFEL, Holzkoppel 4, 22869 Schenefeld, Germany}%

\author{Sergey Usenko}
\affiliation{European XFEL, Holzkoppel 4, 22869 Schenefeld, Germany}%

\author{Zoltan Harman}
\affiliation{Max-Planck-Institut f\"ur Kernphysik, Saupfercheckweg 1, 69117 Heidelberg, Germany}%

\author{Christoph H. Keitel}
\affiliation{Max-Planck-Institut f\"ur Kernphysik, Saupfercheckweg 1, 69117 Heidelberg, Germany}%

\author{Maurice~Leutenegger}
\affiliation{NASA Goddard Space Flight Center, 8800 Greenbelt Rd,
Greenbelt, MD 20771, USA}

\author{Michael Meyer}
\affiliation{European XFEL, Holzkoppel 4, 22869 Schenefeld, Germany}%

\author{Thomas Pfeifer}
\affiliation{Max-Planck-Institut f\"ur Kernphysik, Saupfercheckweg 1, 69117 Heidelberg, Germany}%

\author{Jos\'e R.~{Crespo L\'opez-Urrutia}}
\affiliation{Max-Planck-Institut f\"ur Kernphysik, Saupfercheckweg 1, 69117 Heidelberg, Germany}%

\author{Thomas M.~Baumann}
\email{thomas.baumann@xfel.eu}
\affiliation{European XFEL, Holzkoppel 4, 22869 Schenefeld, Germany}%

\date{\today}

\begin{abstract}

Ultraintense pulses from X-ray free-electron lasers can drive, within femtoseconds, multiple processes in the inner shells of atoms and molecules in all phases of matter.
The ensuing complex ionization pathways of outer-shell electrons from the neutral to the final highly charged states make a comparison with theory enormously difficult.
We resolve these pathways by preparing highly charged ions in an electron beam ion trap before exposing them to the pulsed radiation.
This reveals how relativistic fine-structure effects shift electronic energies, largely compensate the core-screening potential, and enable the consecutive, resonant absorption of two quasi-monochromatic X-ray photons that would generally be unfeasible. This doubly-resonant channel enhances the efficiency of two-photon ionization by more than two orders of magnitude, dominating in this regime the nonlinear interaction of light and matter with possible application for future precision X-ray metrology.

\end{abstract}

\maketitle

Relativistic effects and interelectronic interactions are essential in atomic, molecular, and condensed matter physics. Lasers and high-harmonic generation sources usually probe the dynamics of strongly correlated outer-shell electrons with only small relativistic perturbations, while X-ray free-electron lasers (XFEL) can access the ultrafast dynamics of inner-shell electrons with strong relativistic effects in atoms \cite{young2010,rudek2012,fukuzawa2013,rohringer2012atomic}, molecules \cite{rudenko2017,Kanter2011,erk2014} and metallic foils \cite{vinko2012creation,tamasaku2014x,yoneda2015atomic,mercadier2024transient}. These techniques employ X-rays at photon energies well above the ionization threshold of the inner-shell electrons of various atomic constituents of matter. The high peak fluence of the XFEL pulse opens up ionization trajectories with rapid multiple photoionizations of core electrons, interleaved with autoionization of outer-shell electrons all occurring within a few femtoseconds \cite{sorokin2007,young2010,fukuzawa2013,rorig2023}, 
enabling the generation of atomic X-ray lasers via stimulated emission \cite{rohringer2012atomic,yoneda2015atomic}, the initiation of Coulomb explosion for molecular imaging \cite{erk2014,boll2022x}, and the production of warm-dense plasma of interest in astrophysics research \cite{vinko2012creation,mercadier2024transient}.

Accurate \textit{ab initio} full-relativistic calculations are critical for inner-shell electrons in high-$Z$ elements due to the $\sim Z^{4}$ scaling of the finestructure splitting \cite{gillaspy2001} in order to reproduce the charge state distribution (CSD) beyond the single-photon ionization limit following XFEL irradiation \cite{rudek2013,rudek2018,Toyota2017,Ye2023}. 
Although recent experiments~\cite{rorig2023} have explored the photon-energy dependence of these effects in xenon, extracting the relativistic contributions and predicting the actual ionization trajectories is utterly difficult due to the open shell character of the transient states leading to millions of states which can be populated by the outer electrons and the associated screening induced energy shifts scaling linearly in $Z$.

Here, we observe a strong interplay of relativistic and electron screening effects in highly charged ions (HCIs), enabling enhanced two-photon ionization through resonant double-core hole excitation. We exemplify this in highly charged krypton ions using ultrafast nonlinear spectroscopy \cite{Kanter2011,doumy2011,rudek2012,tamasaku2014x,rudek2018,laforge2021,rorig2023}. 

We expose these ions, produced and trapped in an electron beam ion trap (EBIT \cite{marrs1994}) to ultra-intense pulses from the European X-ray free-electron laser (EuXFEL). With most outer-shell electrons absent, we can investigate this interplaying effects at high resolution, while starting from neutral atoms introduces a ladder of intermediate charge states blurring them. For specific transitions in Kr$^{26+}$, a neon-like ion, we discover that both effects nearly cancel each other out thereby shifting two resonances into the photon-energy bandwidth, resulting in strong multiphoton ionization. 

\begin{figure*}
    \centering
    \includegraphics[width = \textwidth]{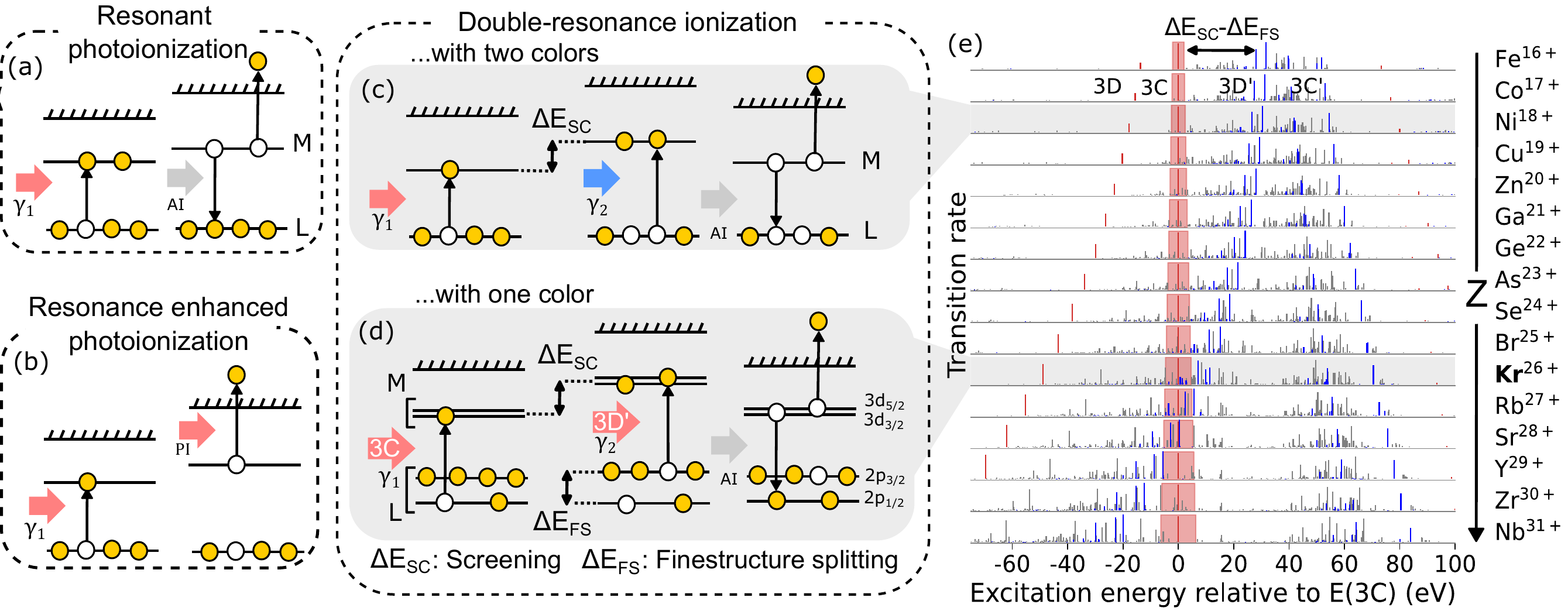}
    \caption{Resonance-enabled X-ray photoionization:
    (a) Linear: Resonant excitation ($\gamma$) followed by autoionization (AI).
(b) Nonlinear REMPI: Resonant excitation followed by direct photoionization.
(c, d) Double-resonance ionization: Two sequential resonances populate a doubly excited autoionizing state.
(e) Calculated absorption spectra for the isoelectronic sequence of ten-electron (Ne-like) ions from iron to neodymium. Red: $3C$ ($2p_{1/2} \rightarrow 3d_{3/2}$) and $3D$ ($2p_{3/2} \rightarrow 3d_{5/2}$) transitions to singly excited states. Grey/blue: Transitions to doubly excited autoionizing states. The $3C$ position is set to zero as a reference. Blue spectra ($3C'$, $3D'$), with a spectator electron, represent absorption after initial $3C$ excitation. For iron, ($3C$, $3D$) and ($3C'$, $3D'$) are well separated due to screening differences ($\Delta \text{E}_{\text{SC}} \approx 40\,\text{eV}$, $\Delta \text{E}_{\text{FS}} \approx 15\,\text{eV}$); in heavier ions, relativistic effects shift $3D'$ toward $3C$ that leads to a total cancellation i.e. $\Delta \text{E}_{\text{SC}}-\text{E}_{\text{FS}} \approx 0$ for Kr, Rb and Sr.
The red band shows the SASE XFEL bandwidth (~0.5\%) near the $3C$ energy, as determined from this experiment—much narrower than the fine-structure splitting in Kr$^{26+}$.}
    \label{fig:REMPIvsDCH}
\end{figure*}

With the XFEL photon energy below the $L$-edge threshold of the krypton ions, ionization of the $L$-shell electron proceeds by resonant excitation shown in Fig.~\ref{fig:REMPIvsDCH}a,b. Here, Fig.\,\ref{fig:REMPIvsDCH}a depicts how this triggers Auger-Meitner autoionization as dominant channel for ions with valence electrons. Meanwhile, Fig.\,\ref{fig:REMPIvsDCH}b illustrates the well-known resonance-enhanced multiphoton ionization (REMPI), a nonlinear process where resonant photoexcitation enables direct photoionization of the excited electron. In both cases, the core vacancy modifies the screening potential such that the photons are no longer in resonance with any transitions of the remaining core. Therefore, resonant excitation of multiple core electrons (shown in Fig.~\ref{fig:REMPIvsDCH}c) at one photon energy is only rarely possible when the bandwidth of the photon source~\cite{pelimanni2024} is very broad or populates close-lying states with high principal quantum number $n$ orbitals~\cite{rorig2023}. 

We exploit instead a nonlinear doubly-resonant two-photon ionization scheme. As shown in Fig.~\ref{fig:REMPIvsDCH}d, relativistic effects split the $2p$ subshell into $2p_{1\!/\!2}$ and $2p_{3\!/\!2}$ orbitals, with the subscript being the single-electron total angular momentum of each orbital. While the first photon resonantly excites a $2p_{1\!/\!2}$ electron, the second photon (with the same photon energy) can now resonantly excite a $2p_{3\!/\!2}$ electron -- when the change in the core-screening potential is compensated by the relativistic fine-structure splitting. Then, a fast Auger-Meitner autoionization of these doubly excited state efficiently generates higher charge states. The rate enabled by this double-resonance is orders of magnitude faster than that of the conventional single-resonance REMPI scheme, thus driving far stronger nonlinear ionization.  
\begin{figure}
    \centering
    \includegraphics[width=0.5\textwidth]{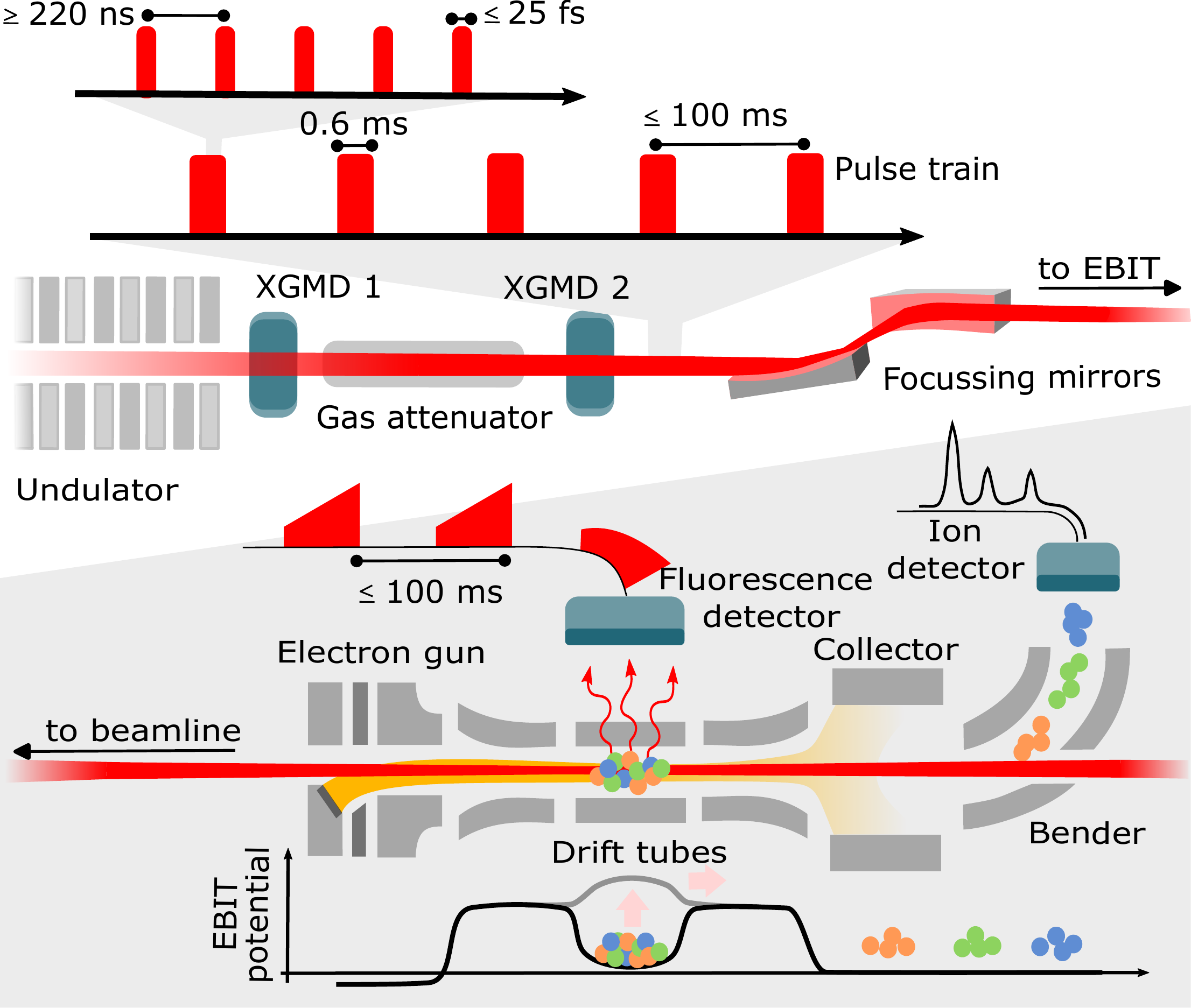}
    \caption{Top: Scheme of the SASE3 beamline at EuXFEL. XFEL pulses with the depicted pattern from an undulator pass XGMD1 (for intensity measurements), a gas attenuator and a second intensity monitor (XGMD2). Two KB mirrors focus the beam into the experimental chamber (bottom) of SQS-EBIT to irradiate HCIs produced and trapped in it. A silicon drift detector records their fluorescence, while an ion-ToF spectrometer downstream of the EBIT registers their charge-state distribution after each exposure period. The bottom panel depicts the electrostatic potential of EBIT and ion extraction beamline.}
    \label{fig:AufbauExp}
\end{figure} 
\begin{figure*}
    \centering
    \includegraphics[width=\textwidth]{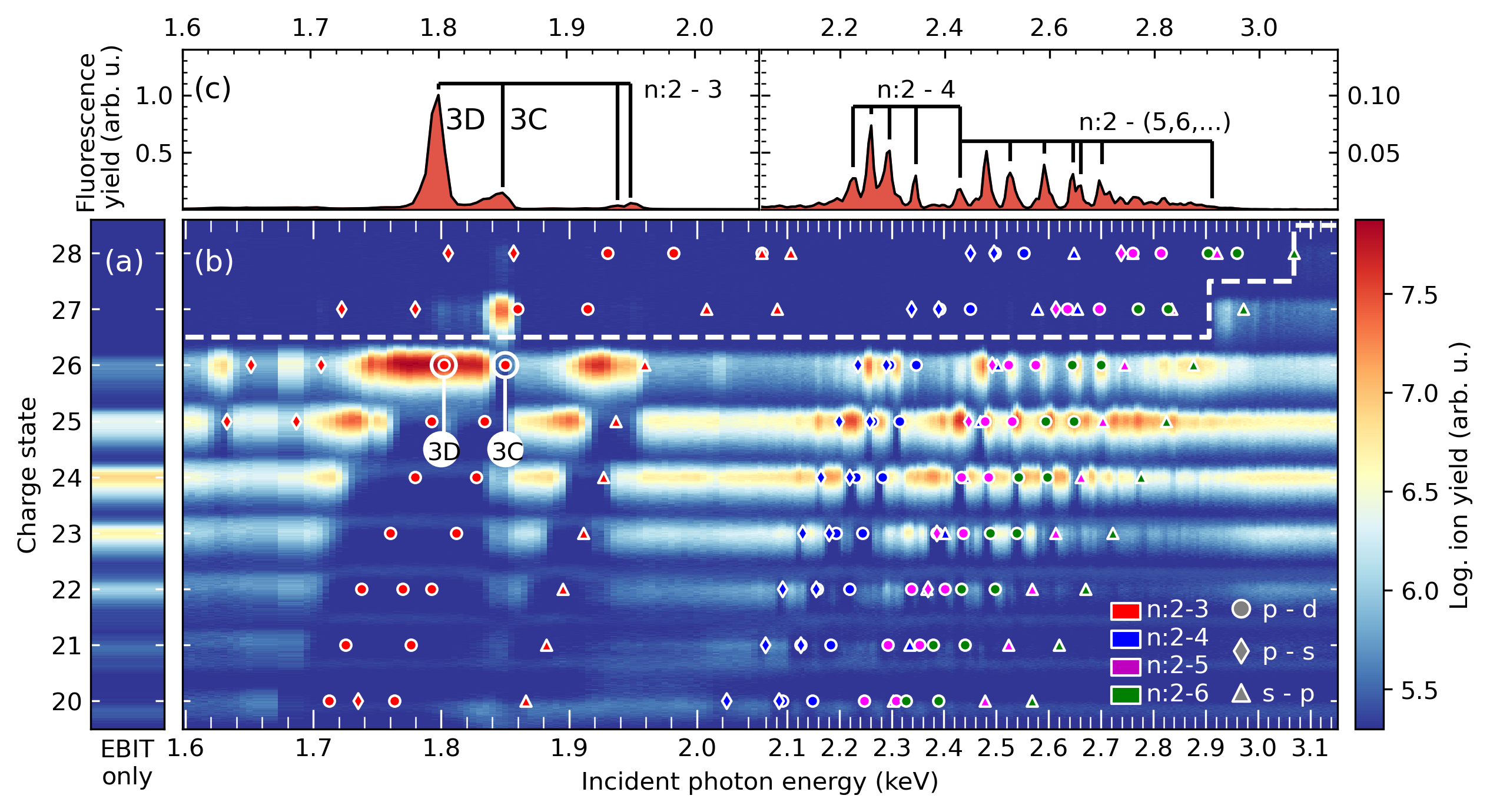}
    \caption{Overview spectrum from 1.6\,keV to 3.1\,keV with highly charged krypton in EBIT. (Top) Fluorescence yield; (bottom) charge-state distribution (CSD) versus incident photon energy; (left) initial CSD without XFEL irradiation. Note the change in the scaling of the photon-energy axis around 2.1\,keV. Predicted energies of major ground-state transitions for each charge state are marked with colors according to the upper state $n$  and symbols for the orbital angular momentum, e.g., blue circles for $2p\rightarrow4d$ transitions. The white dashed line marks the ionization threshold.}
    \label{fig:Overviewspectrum}
\end{figure*}

Our \textit{ab initio} relativistic atomic-structure calculations using the Flexible Atomic Code (FAC)\,\cite{gu2008} predict that such double resonance phenomena would emerge in mid-heavy elements, where relativistic fine-structure effects
can counterbalance Coulomb screening. In Fig.~\ref{fig:REMPIvsDCH}e, we show the calculated absorption spectra of the $2p\rightarrow3d$ transition in neon-like isoelectronic HCIs.
In such systems, there are very strong, well-known transitions dubbed $3C$ and $3D$:
\begin{equation}
    3C : \left[ 2p^6 \right]_{J=0} \rightarrow \left( \left[ 2p^5 \right]_{1/2} 3d_{3/2} \right)_{J=1} \label{eq.3C}
\end{equation} 
and
\begin{equation}
        3D : \left[ 2p^6 \right]_{J=0} \rightarrow \left( \left[ 2p^5 \right]_{3/2} 3d_{5/2} \right)_{J=1}
\end{equation}
following the notation of \citet{gabriel1972}. 
In such a notation, $3C$ represents the $2p_{1\!/\!2}\rightarrow3d_{3\!/\!2}$ transition and $3D$ is the $2p_{3\!/\!2}\rightarrow3d_{5\!/\!2}$ transition. 
Their positions and strengths are indicated by the red lines. 
We choose the position of the $3C$ resonance in each element as zero for referencing the other transitions. Grey and blue lines show the absorption spectra for a second photon after the HCIs are excited by the first one (i.e., the $2p^53d\rightarrow2p^43d^2$ transition). Particularly, blue lines highlight the second resonant excitations arrays marked as $3D'$ and $3C'$ following a resonant excitation of $3C$ (see Eq.~(\ref{eq.3C})).
Two distinct scaling laws appear clearly in the spectra in Fig.\,\ref{fig:REMPIvsDCH}e: 
firstly, the relativistic splitting between the $2p_{1\!/\!2}$ and $2p_{3\!/\!2}$ orbitals scaling with $Z^4$ for the $3D$ and $3D'$ transitions, and secondly, the changing screening potential due to the extra hole state in the $2p$ subshell that scales linearly in $Z$ for the $3C'$ transitions with respect to $3C$. Within this $Z$ range, the splitting $3d_{3\!/\!2}$--$3d_{5\!/\!2}$ is negligible by comparison.
In lighter elements with weak relativistic effects in the $2p$ subshell, the strong screening potential pushes the second excitation $\gtrsim30$~eV above the first one, making it within an XFEL bandwidth of approximately $10$~eV impossible to hit the second resonance transition at once \cite{bernitt2012}.
However, with growing $Z$, the $Z^4$ scaling relativistic splitting brings both the $3C$ and $3D'$ transitions for Kr, Rb, Sr, and Y within the bandwidth of the XFEL pulses, enabling a second resonant excitation. 
\begin{figure*}
    \centering
    \includegraphics[width=\textwidth]{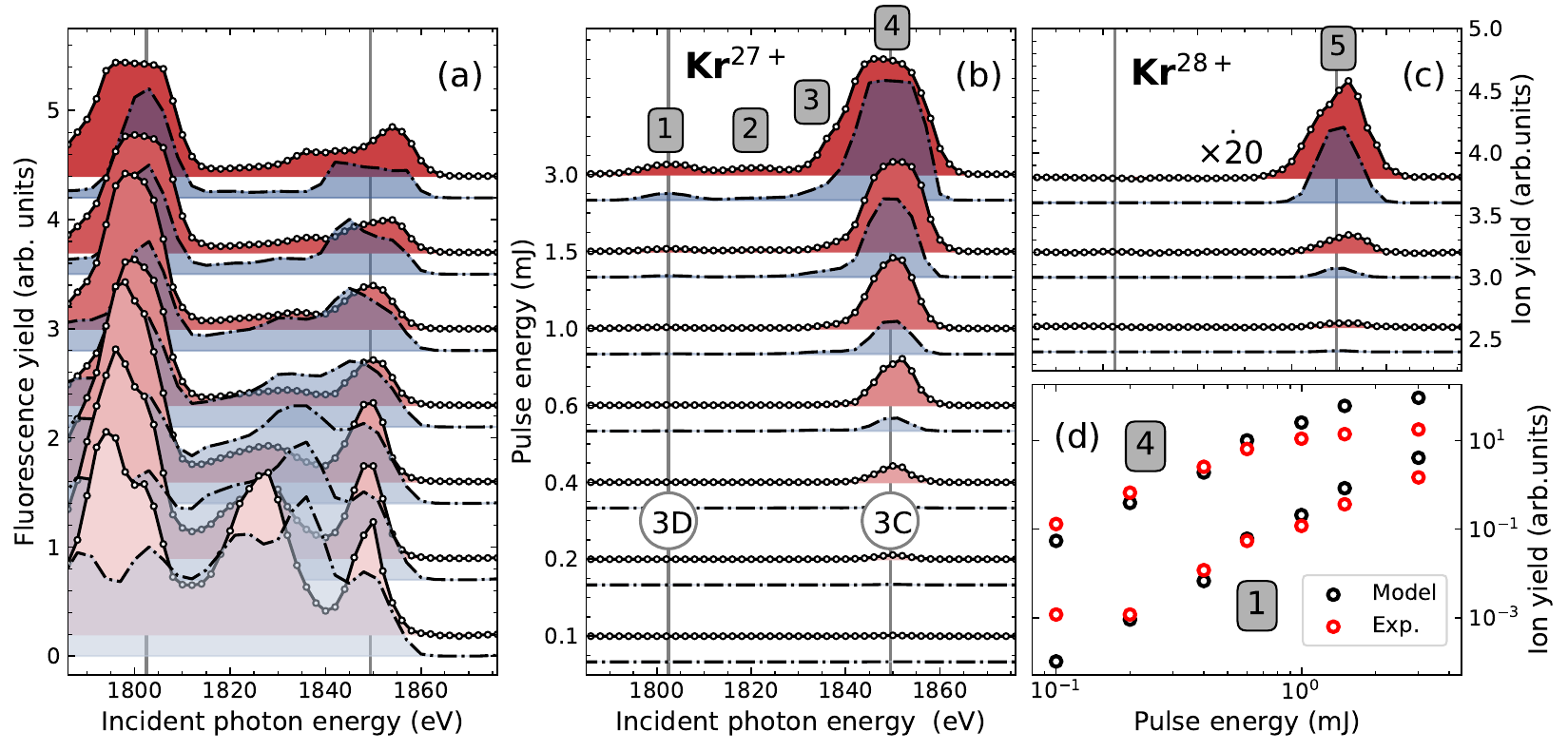}
    \caption{Experimental data (red) and predictions (grey) for the photon energies near the $3C$ and $3D$ resonant transitions for varying pulse energy. (a): Average fluorescence yield. (b): ion-yield spectra for fluorine-like krypton. (c): ion yield spectra for oxygen-like krypton. Five identified peaks are labeled from 1 to 5. (d) Comparison of Kr$^{27+}$ yield evaluated at 1800\,eV (peak (1)) and 1850\,eV (peak (4)) as a function of the pulse energy.}
    \label{fig:KryptonOverview}
\end{figure*}

To study this, we deploy an electron beam ion trap at the Small Quantum Systems (SQS) instrument at the SASE3 soft X-ray undulator of the European X-ray Free-Electron Laser facility (see Fig.\,\ref{fig:AufbauExp}). The so-called SQS-EBIT (based on \cite{micke2018}) prepares a dilute cloud of trapped highly charged ions with a charge-state distribution (CSD) peaking for the present experiment at 36\% Kr$^{24+}$ and with approximately 4\% neon-like Kr$^{26+}$ (Fig.\,\ref{fig:Overviewspectrum}a, for more details see End Matter).
While maintaining ultrahigh fluences, EuXFEL is tuned to emit at a bandwidth of approximately 0.5\%, significantly narrower than the typical 1\% bandwidth produced by SASE XFELs. This allows for a more selective excitation.
We scan the X-ray photon energy from 1.6\,keV up to 3.2\,keV to cover all the possible resonant transitions of the $L$-shell electrons of Kr ions. XFEL radiation is generated in bursts at a repetition rate of 10\,Hz, each one containing 500 pulses with a maximum pulse energy of about 3\,mJ at 1.8\,keV, decreasing steadily to about 1\,mJ at 3\,keV. To obtain the average CSD in Fig.\,\ref{fig:Overviewspectrum}b at each photon energy, SQS-EBIT accumulates photoions during 5000 XFEL pulses and is emptied at 1\,Hz. We also record the corresponding integrated fluorescence flux from all trapped HCIs (see Fig.\,\ref{fig:Overviewspectrum}c).

Given the predicted ionization potentials of Kr$^{26+}$ and Kr$^{27+}$, 2929\,(2) and 3072\,(5)\,eV \cite{biemont1999}, respectively, the white dashed line in Fig.~\ref{fig:Overviewspectrum}b indicates the limit for charge states accessible by direct photoionization; below it, $M$-shell electrons can be removed and the CSD modified. However, the comparison between Figs.~\ref{fig:Overviewspectrum}a and~\ref{fig:Overviewspectrum}b shows that such processes have a rather minute effect on the CSD. Only resonant excitation by XFEL photons can efficiently bring the ions to higher charge states. Depletion and accumulation of the different charge states appear at the resonant transitions of $L$-shell electrons to $n\,\geq\,3$ shells predicted with our atomic structure code. The excited electron can either be brought into the continuum by absorbing another photon by a nonlinear REMPI process (Fig.\,\ref{fig:REMPIvsDCH}b), or trigger Auger-Meitner autoionization of other $M$-shell electrons (Fig.\,\ref{fig:REMPIvsDCH}a). Because this latter linear process ionizes more efficiently, the dark and bright spots observed in Fig. \ref{fig:Overviewspectrum}b are mainly caused by it. 

Signatures of nonlinear multiphoton ionization processes become much more clear for charge states above the direct photoionization limit shown in Fig.\,\ref{fig:Overviewspectrum}. The lack of $M$-shell electrons in the ground state for ions above Kr$^{26+}$ excludes linear one-photon ionization and one-photon excitation--autoionization processes in this region. 
Consequently, the predicted $3C3D'$ double-resonance two-photon ionization appears at photon energies around 1850\,eV with a much enhanced production of Kr$^{27+}$ ions.

To investigate these nonlinear processes in more detail, we show the  photoion spectra of Kr$^{27+}$ and Kr$^{28+}$ covering the 3C and 3D excitation energies (1.79-1.88\,keV) of Kr$^{26+}$ at pulse energies between 3\,mJ and 0.1\,mJ (red spectra in Fig.\,\ref{fig:KryptonOverview}). The photon yield that is additionally recorded is shown as  fluorescence
spectra, serving as auxiliary references.
Furthermore, we solve a set of rate equations to model the interaction of trapped ions with one entire XFEL pulse train to simulate the spectra (grey spectra in Fig.\,\ref{fig:KryptonOverview}, see Methods for further details).
In general, we find good agreement between the recorded and simulated spectra. Especially, the ion-yield spectra closely resemble the distinct predictions of the two nonlinear ionization processes introduced in Figs.\,\ref{fig:REMPIvsDCH}b and \ref{fig:REMPIvsDCH}d. 
I.e., following the excitation of the neon-like ion with the $3D$ transition at 1800\,eV by the first photon, absorption of a second photon is energetically prohibited due to the modification of the screening potential by the $2p_{3\!/\!2}^{-\!1}3d$ hole.
Therefore, production of Kr$^{27+}$ can only proceed by slow photoionization of the excited $3d$ electron in Kr$^{26+}$ (Peak (1) in Fig.\,\ref{fig:KryptonOverview}b), which can also decay back to the ground state by emitting fluorescent photons. This results in the fluorescence peak around 1800\,eV shown in Fig.\,\ref{fig:KryptonOverview}a. On the other hand, tuning the photon energy to the $3C3D'$ double resonances around 1850\,eV yields a large fraction of Kr$^{27+}$ ions (Peak (4) in Fig.\,\ref{fig:KryptonOverview}b) by fast autoionization of the doubly excited $2p_{1\!/\!2}^{-\!1}2p_{3\!/\!2}^{-\!1}3d^2$ state in Kr$^{26+}$, also suppressing fluorescence in Fig.\,\ref{fig:KryptonOverview}a. This signatures confirm the double-resonance degeneracy predicted for neon-like krypton illustrated in Fig.\,\ref{fig:REMPIvsDCH}. The discrepancy between the observed and simulated fluorescence spectra at low pulse intensities may be attributed to the angular distribution of the fluorescence, which is not properly included in the model.

Furthermore, both Kr$^{27+}$ and singly-excited Kr$^{26+}$ have a hole in the core (L-shell) causing similar screening such that the $3D$ transition in Kr$^{27+}$ is close to the $3C3D'$ double resonance transitions in neon-like Kr$^{26+}$. This opens an efficient pathway for production of Kr$^{28+}$ at a photon energy around 1850\,eV, comprising in total four-photon sequential ionization from the ground state of Kr$^{26+}$ as shown Fig.\,\ref{fig:KryptonOverview}. 
{The relatively narrow photonenergy bandwidth of approximately 0.5\% improves the selectivity, thereby enabling this excitation process for only a few elements along the isoelectronic sequence, where the 3C-3D' consecutive resonant excitation condition is fulfilled. Furthermore, the enhanced resolution reveals additional ionization channels that are initiated from lower charge states. These are observed as extra peaks at 1820\,eV (Peak\,(2)) and 1840\,eV (Peak\,(3)) in the photoion spectra of Kr$^{27+}$ (Fig.\,\ref{fig:KryptonOverview}b). They correspond to analogous multiphoton ionization pathways mediated by doubly excited states in Kr$^{24+}$ and Kr$^{25+}$, respectively. 
This shows that blends of single and double excitation (see Fig.\,\ref{fig:REMPIvsDCH}) also occur for charge states below neon-like krypton. While the double-resonance ionization pathway easily reaches saturation, we infer from the extracted ratio between the Kr$^{27+}$ ion yields seen in Fig.\,\ref{fig:KryptonOverview}d at photon energies of 1800 and 1850\,eV that the double-resonance ionization pathway is more than two orders of magnitude more efficient than the single-resonance ionization pathway, which is in good agreement with the model. Slight discrepancies such as in the reproduction of the depletion trend at high intensities in Fig.\,\ref{fig:KryptonOverview}d are tentatively attributed to the model limitations. In particular, the model approximates unknown parameters such as the SASE spectral, temporal characteristics and intensity distribution at the focus (see End Matter for details).

In conclusion, we have predicted and experimentally characterized ionization channels with sequential nearly degenerate resonant core-excitations forming doubly excited states. 
For this, we performed ultraintense nonlinear spectroscopy on highly charged ions produced and confined in an electron beam ion trap. So far in the context of ultraintense-light matter interaction, such a mechanism was only considered feasible in rare cases where the bandwidth of the exciting photon beam was broad enough. Although with single-color X-ray lasers, this nonlinear process is only accessible for particular elements where relativistic effects compensate changes in electronic screening potentials, it can be generalized to most systems by employing two-color X-ray pulses, also available at EuXFEL \cite{Serkez2020} (see Fig.\,\ref{fig:REMPIvsDCH}c). 
Therefore, our work paves the way for efficient state-selective two-color pump-probe scheme through multiply excited autoionizing states, which can be applied to future precision tests of physics as well as x-ray optical clocks based on envisioned XFELO- and frequency-comb x-ray sources.


\section{Acknowledgement}
We acknowledge European XFEL in Schenefeld, Germany,
for provision of beam time at the SQS instrument and would like to thank the staff for their skillfull support. We thank the fine-mechanics shop lead by Thorsten Spranz at the Max-Planck-Institute for Nuclear Physics for manufacturing the SQS-EBIT. Research was funded by the Max-Planck-Gesellschaft
(MPG), Germany. C.S. acknowledges support from MPG and NASA-JHU
Cooperative Agreement. 
KK and MS have been funded by the Deutsche Forschungsgemeinschaft (DFG, German Research Foundation) – 497431350.

Data recorded for this experiment at the European XFEL are available at doi:10.22003/XFEL.EU-DATA-003315-00.

\section{End matter}

\subsection{Experimental set up}
SQS-EBIT follows the design of the Heidelberg Compact EBIT \cite{micke2018} and is equipped with an off-axis electron gun.
Passing through a two-stage differentially pumped injection system, monoisotopic krypton gas is led into the interaction zone, where SQS-EBIT produces a magnetically focused electron beam of approximately 10\,mA and 1780\,eV, well below the ionization threshold for neon-like krypton.
Then, efficient electron impact ionization drives up the center of gravity of the charge-state distribution of krypton ions. After it equilibrates with counteractive recombination processes, we find a charge-state distribution of highly charged ions showing a maximum abundance of 36\% magnesium-like Kr$^{24+}$ ions and approximately 4\% neon-like Kr$^{26+}$ (See left panel in Fig.\,\ref{fig:KryptonOverview}). All ions are trapped by an axial electrostatic trapping potential of approximately 10\,V and by the negative space charge of the electron beam. 

The MHz-pulse-train of European XFEL is introduced to SQS-EBIT through an aperture in its off-axis electron gun and illuminates the ion cloud, before leaving the instrument unimpeded. The ultrashort pulses, with estimated pulse duration of about 30\,fs, are focused onto the ion cloud by means of Kirkpatrick- Baez (KB) optics, producing an estimated focal radius of 10\,$\mu$m. The interaction of the X-ray pulse with the trapped ion cloud is monitored by a silicon drift detector (SDD) directed towards the trapped ions in a plane perpendicular to the linear polarization of the incoming XFEL radiation. Fluorescence photons are recorded in coincidence with the incoming train of XFEL pulses. The charge-state distribution shortly after the ion cloud has interacted with the XFEL trains is recorded by ejecting the trapped ions into a time-of-flight spectrometer. After a time-of-flight of approximately 1\,$\mu s$, charge-state separated ions are collected by an multichannel plate in current mode. In Fig.\ref{fig:Overviewspectrum}, the time-of-flight axis is transformed to present all charge states equidistantly.

\subsection{Simulation}
To quantitatively interpret and analyze the dependence of the line ratio under the influence of resonant doubly excited states, we have simulated the XFEL-HCI interaction using a simple model based on a set of rate equations.
For this purpose, we generate the necessary atomic data for the interactions for oxygen-like down to silicon-like krypton by utilizing
the flexible atomic code (FAC\,\cite{gu2008}) in the UTA-mode (unresolved transition array mode). In particular, we include all singly $1s^2\,2l^{-1}\,3l'^1$ and doubly core-excited $1s^2\,2l^{-2}\,3l'^2$ states. Additional higher-excitations are included to improve precision due to configuration interaction. We take into account absorption ($B_{ij}$), stimulated emission ($B_{ji}$) that connects the ground state with singly excited and singly excited with doubly excited states.
Furthermore, we predict radiative decay ($A_{ji}$), photoionization ($\sigma_{ij}$) as well as Auger-Meitner decays ($A^{\text{AI}}_{ji}$) to all possible levels.
Slow processes such as interactions involving the electron beam of SQS-EBIT as well as recombination with surrounding residual and injected gas are neglected (for 
more details refer to next chapter). 
Non-sequential processes are not included.
Furthermore, we verified the energies as well as absorption rates with calculations using the GRASP code.\\
With all the processes calculated, we solve the well-known set of coupled rate equations \cite{young2010}
\begin{equation}
    \frac{d}{dt} P_i(t) = \sum_{j \neq i}^{\text{all levels}} \left[ \Gamma_{j \to i} P_{j}(t) - \Gamma_{i \to j} P_i(t) \right]
    \label{eq1}
\end{equation}
to predict the time evolution of the populations P$_{i}$ of all levels $i$. The rate $\Gamma_{i \to j}$ corresponds to the transition probability between levels $i$ and $j$.
In order to relate the intensity-dependent photoionization crosssections and Einstein B coefficients to the radiative and Auger-Meitner decays, we use the following estimated parameters:
$\sigma$\,(\text{XFEL focus, FWHM}) = 10\,$\mu$m, $\tau$\,(\text{pulse duration}) = 30\,fs, E$_{\text{max}}$\,(\text{max. pulse energy}) = 3\,mJ. For the photon energy bandwidth, we find a value of 0.5\,\% of the photon energy to be the most suitable. We further assume the spectral energy distribution to be Gaussian, neglecting any spectral properties arising from the SASE process. Additionally, we approximate the temporal profile as a square pulse of duration of $\tau$.\\
Under these approximations, we solve Eq.\,\ref{eq1}  and predict the time evolution of the involved level populations during and after interaction with the XFEL pulse. 
We expand the model to include the train structure of the incoming XFEL pulses (500 pulses separated at 880\,ns) by performing the one-pulse simulation iteratively using the output as input for the next. Furthermore, the photon energy is scanned for comparison with the experiment.
With this model, the dominant ionization channels producing peak (1) and (4) in the photoion spectrum of fluorine-like krypton can be identified as:
\begin{equation}\nonumber
    \textbf{(P1)}: \,\, \text{Kr}^{26+} \xrightarrow{\text{R}} (\text{Kr}^{26+})^* \xrightarrow{\text{PI}} \text{Kr}^{27+}
\end{equation}
and peak (4):
\begin{equation}\nonumber
    \textbf{(P4)}:\,\, \text{Kr}^{26+} \xrightarrow{\text{R}} (\text{Kr}^{26+})^{*} \xrightarrow{\text{R}}  (\text{Kr}^{26+})^{**} \nonumber\\ \xrightarrow{\text{AI}} \text{Kr}^{27+}.
\end{equation}
The asterisk denotes the number of excitations.
Both are two-photon processes, however peak (4) is associated with a significantly higher cross section, which becomes apparent due to its fully-resonant property.
Peak (2) is initiated by a resonance from the ground state of Kr$^{24+}$ (Mg-like Kr):
\begin{equation}\nonumber
\begin{aligned}
\label{Eq.: Scattering matrix element}
        \textbf{(P2)}:\,\, \text{Kr}^{24+} \xrightarrow{\text{R}}
        (\text{Kr}^{24+})^*
        \xrightarrow{\text{R}} (\text{Kr}^{24+})^{**}\xrightarrow{\text{AI}} (\text{Kr}^{25+})^* \nonumber\\\xrightarrow{\text{R}} (\text{Kr}^{25+})^{**}  \xrightarrow{\text{AI}} (\text{Kr}^{26+})^*\xrightarrow{\text{PI}} \text{Kr}^{27+}
\end{aligned}
\end{equation}
Peak (3) is initiated by a resonance from the ground state of Kr$^{25+}$ (Na-like Kr):
\begin{equation}\nonumber
\begin{aligned}
\label{Eq.: Scattering matrix element}
       \textbf{(P3)}:\,\, \text{Kr}^{25+} \xrightarrow{\text{R}} (\text{Kr}^{25+})^* \xrightarrow{\text{R}} (\text{Kr}^{25+})^{**} \nonumber \\ \xrightarrow{\text{AI}} \text{Kr}^{26+} \xrightarrow{\text{R}} \text{Kr}^{26+}
        \xrightarrow{\text{AI}} \text{Kr}^{27+}
\end{aligned}
\end{equation}
Spectral feature E is identified as a continuation of D with:
\begin{equation}\nonumber
    \textbf{(P5)}:\,\, \text{[(P4)]} \xrightarrow{\text{R}} (\text{Kr}^{27+})^* \xrightarrow{\text{PI}} \text{Kr}^{28+},
\end{equation}
which makes peak (5) with its entire process from the ground state neon-like krypton to oxygen-like krypton, effectively a sequential four photon process, peak (2) and (3) a sequential six and four photon process, respectively. \\
The simulation does not take into account the fluence distribution at the XFEL focus nor do we take into account the spatial distribution of the trapped HCI cloud, typically assumed to be cigar shaped along the (electron) beam propagation direction.
However, the good qualitative agreement between the data and model, along with the reasonable estimates for the unknown XFEL parameters, places a more rigorous treatment of the HCI-XFEL interaction outside the scope of this work.

\end{document}